\documentclass[12pt]{article}

\begin{document}

\title{Extra symmetry in the field equations in $5D$ with spatial spherical symmetry}
\author{J. Ponce de Leon\thanks{E-mail: jponce@upracd.upr.clu.edu, jpdel1@hotmail.com}\\ Laboratory of Theoretical Physics, Department of Physics\\ 
University of Puerto Rico, P.O. Box 23343, San Juan, \\ PR 00931, USA} 
\date{February  2006}

\maketitle

\begin{abstract}
We point out  that the field equations in $5D$, with spatial spherical symmetry, possess an extra symmetry that leaves them invariant. This symmetry corresponds to certain simultaneous interchange of coordinates and  metric coefficients. As a consequence a single solution in $5D$   can generate very different scenarios in $4D$, ranging from static configurations to cosmological situations. A new perspective emanates from our work. Namely, that different astrophysical and cosmological scenarios in $4D$ might correspond to the same physics in $5D$. We present explicit examples that illustrate this point of view.

\end{abstract}

\medskip

PACS: 04.50.+h; 04.20.Cv

{\em Keywords:} Kaluza-Klein Theory; Brane Theory; Space-Time-Matter Theory; General Relativity. 

\newpage
\section{Introduction}
Most of the recent advances in theoretical physics deal with models of our universe in more than four dimensions.  Theories of the Kaluza-Klein type in many dimensions are used in different branches of physics. Superstrings $(10D)$ and supergravity $(11D)$ are well known examples. In gravitation and cosmology  braneworld models \cite{Arkani1}-\cite{Maartens2} in $5D$ as well as space-time-matter (STM) theory  \cite{Wesson book} have become quite popular. 

All these theories face the same challenge, namely the prediction of   observable effects from the extra dimensions \cite{JpdeLgr-qc/0111011}. The success of this mission depends on the ``correct" identification of the physical or observable space-time metric from the multidimensional one. This is not a trivial task in higher-dimensional cosmologies, like  braneworld and STM, where the extra dimension
 is non-compact. 
In this scenario all the coordinates are alike, in the sense that the metric tensor is allowed to depend explicitly on the extra coordinate, usually called {\it fifth} coordinate. 

In this regard, the crucial question is: given an arbitrary $5$-dimensional metric, that depends on all five coordinates, how do we decide which one is the ``extra" coordinate?. The answer to this question seems to be far from obvious. Even in the simple case of spherical symmetry, in ordinary three space, there are various possible options leading to different scenarios in $4D$.

In  the present paper we point out  that the field equations in $5D$, with spatial spherical symmetry, possess an extra symmetry that leaves them invariant. This extra symmetry corresponds to certain simultaneous interchange of coordinates and  metric coefficients. As a consequence, for every solution of the field equations in $5D$, there are different options for the identification of the extra dimension. 

Our main conclusion is that, by virtue of the abovementioned extra symmetry, a single solution in $5D$   can generate very different scenarios in $4D$, ranging from static configurations to cosmological situations. Said another way, the additional symmetry allows the unification in $5D$ of diverse, {\it apparently} different,  physical scenarios in $4D$.

The paper is organized as follows. In section $2$ we give a brief summary of  the general splitting formalism in $5D$ which leads to an effective, or induced, energy-momentum in $4D$. In section $3$ we discuss the stated additional symmetry  of the field equations for the case of spatial spherical symmetry.  

In section $4$, in the scenario where the metric is independent on the fifth coordinate, we show that the extra  coordinate  can be either spacelike or timelike, without affecting the interpretation in $4D$. At this point it is worth to mention that  physical conditions, imposed on the $4D$ effective matter, 
do not preclude the existence of a {\it large} timelike extra dimension\footnote{For a critical review of some objections commonly raised against the timelike signature of the extra coordinate see Ref. \cite{JPdeLgr-qc/0212058}}.  Then, we generate ``new" solutions by the appropriate interchange of coordinates and metric coefficients.  
We illustrate the discussion with a specific solution of the $5D$ field equations that produces the following scenarios in $4D$: (i) static configurations, (ii) evolution of inhomogeneities, and (iii) spatially-homogeneous cosmological models of the Kantowski-Sachs type.

In section $5$ we consider the case of $5D$ metrics with explicit dependence on the extra dimension and which allow both signatures. We show that the various  options for the identification of the extra dimension correspond to different looking metrics in $5D$ with distinct interpretation in $4D$.  

\section{Field equations}
In order to make the paper self-consistent and set the notation, in this section we give a brief review  of the formalism in $5D$. The line element is given by
\begin{equation}
\label{5D line element}
dS^2 = \gamma_{AB}dx^Adx^B,
\end{equation}
where $A, B = (0-4)$ and $\gamma_{AB}$  is allowed to depend on all five coordinates. The dynamics in $5D$ is assumed to be governed by the Einstein equations
\begin{equation}
\label{field equations in 5D}
{^{(5)}G}_{AB} = R_{AB} - \frac{1}{2}\gamma_{AB}R = k_{(5)}^2{^{(5)}T}_{AB},
\end{equation}
where $k_{(5)}^2$ is a constant introduced for dimensional considerations and ${^{(5)}T}_{AB}$ is the five-dimensional energy-momentum tensor.

Our $4D$ spacetime is orthogonal to the extra dimension. The unit ($n_{A}n^{A} = \epsilon$) vector, along the fifth dimension 
is given by\footnote{We note that this vector is {\it not} orthogonal to the hypersurfaces $y = const.$, except in the case where $\gamma_{4\mu} = 0$. }   
\begin{equation}
\label{vector along the extra dimension}
n^A = \frac{\delta^{A}_{4}}{\sqrt{\epsilon \gamma_{44}}}\;  , \;\;\;\;\;  n_{A}= \frac{\gamma_{A4}}{\sqrt{\epsilon \gamma_{44}}}, 
\end{equation}
where $\epsilon = -1$ or $\epsilon = + 1$ depending on whether the extra dimension 
is spacelike or timelike, respectively. The metric induced in $4D$ is given by $g_{AB} = \gamma_{AB} - n_{A}n_{B}$. Since $g_{44} = 0$, denoting $\gamma_{44} = \epsilon \Phi^2$ and $\gamma_{\mu 4} = \epsilon \Phi^2 A_{\mu}$, the line element (\ref{5D line element}) can be written as 
\begin{equation}
dS^2 = g_{\mu\nu}dx^{\mu}dx^{\nu} + \epsilon \Phi^2(A_{\mu}dx^{\mu} + dx^4)^2.
\end{equation}
In absence of off-diagonal terms $(\gamma_{4\mu} = 0)$, the vector $n^A$ defined in (\ref{vector along the extra dimension}) becomes orthogonal to hypersurfaces $y = const.$, which provides a clean separation between the spacetime section and the extra coordinate, viz., 
\begin{equation}
\label{metric with g44 not 1}
d{\cal S}^2 = g_{\mu\nu}(x^{\rho}, y)dx^{\mu}dx^{\nu} + \epsilon \Phi^2(x^{\rho}, y) dy^2,
\end{equation}
here and in what follows the extra coordinate will be denoted as $y \equiv x^4$. In this case the dimensional 
reduction of the five-dimensional equations  is particularly simple \cite{EMT}. In fact, using the Gauss-Codacci-Mainardi relations, the $15$ equations (\ref{field equations in 5D}) can be split up into 
three parts. 

The first part consists of $10$ equations which are interpreted as the effective field 
equations in $4D$. They are
\begin{eqnarray}
\label{4D Einstein with T and K}
{^{(4)}G}_{\alpha\beta} &=& \frac{2}{3}k_{(5)}^2\left[^{(5)}T_{\alpha\beta} + 
(^{(5)}T^{4}_{4} - \frac{1}{4}{^{(5)}T})g_{\alpha\beta}\right] -\nonumber \\
& &\epsilon\left(K_{\alpha\lambda}K^{\lambda}_{\beta} - 
K_{\lambda}^{\lambda}K_{\alpha\beta}\right) + 
\frac{\epsilon}{2} g_{\alpha\beta}\left(K_{\lambda\rho}K^{\lambda\rho} - 
(K^{\lambda}_{\lambda})^2 \right) - \epsilon E_{\alpha\beta}, 
\end{eqnarray}
where $K_{\mu\nu}$ is the extrinsic curvature 
\begin{equation}
\label{extrinsic curvature}
K_{\alpha\beta} = \frac{1}{2}{\cal{L}}_{n}g_{\alpha\beta} = \frac{1}{2\Phi}\frac{\partial{g_{\alpha\beta}}}{\partial y},\;\;\; K_{A4} = 0,
\end{equation}
and $E_{\mu\nu}$ is the projection of the bulk Weyl tensor ${^{(5)}C}_{ABCD}$ orthogonal to ${{n}}^A$, i.e., ``parallel" to spacetime, viz.,
\begin{eqnarray}
\label{Weyl Tensor}
E_{\alpha\beta} &=& {^{(5)}C}_{\alpha A \beta B}n^An^B\nonumber \\
&=& - \frac{1}{\Phi}\frac{\partial K_{\alpha\beta}}{\partial y} + K_{\alpha\rho}K^{\rho}_{\beta} - \epsilon \frac{\Phi_{\alpha;\beta}}{\Phi} - \epsilon \frac{k^{2}_{(5)}}{3}\left[{^{(5)}T}_{\alpha\beta} + ({^{(5)}T}^{4}_{4} - \frac{1}{2}{^{(5)}T})g_{\alpha\beta}\right].
\end{eqnarray}
The second part is an 
inhomogeneous wave equation for $\Phi$, which follows from the fact that $E_{\mu\nu}$ is traceless. Indeed, the requirement $E_{\mu}^{\mu} = 0$ is equivalent to ${^{(5)}G}_{44} = k^{2}_{(5)}{^{(5)}T_{44}}$ 
from (\ref{field equations in 5D}) and gives 
\begin{equation}
\label{equation for Phi}
{\Phi}^{\mu}_{;\mu} = - \epsilon \frac{\partial K}{\partial y}-
 \Phi (\epsilon K_{\lambda \rho} K^{\lambda \rho} + {^{(5)}R}^{4}_{4}).
\end{equation}
Finally, the remaining four equations are
\begin{equation}
\label{conservation equation}
D_{\mu}\left(K^{\mu}_{\alpha} - 
\delta^{\mu}_{\alpha}K^{\lambda}_{\lambda}\right) = 
k_{(5)}^2 \frac{{^{(5)}T_{4\alpha}}}{\Phi}.
\end{equation}
In the above expressions, the covariant  derivatives are 
calculated with respect to $g_{\alpha\beta}$, i.e., $Dg_{\alpha\beta} = 0$.

\subsection{Effective energy-momentum tensor in $4D$}
Following the usual procedure, we define the energy-momentum tensor in four-dimensions (on the hypersurface $y = const$) through (\ref{4D Einstein with T and K}), the effective Einstein field equations in $4D$, namely 
\begin{equation}
8 \pi GT_{\mu\nu}^{(eff)} \equiv  - \epsilon\left(K_{\mu\lambda}K^{\lambda}_{\nu} - K_{\lambda}^{\lambda}K_{\mu\nu}\right) + \frac{\epsilon}{2} g_{\mu\nu}\left(K_{\lambda\rho}K^{\lambda\rho} - (K^{\lambda}_{\lambda})^2 \right) - \epsilon E_{\mu\nu}, 
\end{equation}
which, in terms of the metric, is given by (in what follows $\stackrel{\ast}{f} \equiv \partial f/\partial y$), 

\begin {eqnarray}
\label{EMT in STM}
8 \pi GT_{\mu\nu}^{(eff)} =  &-& \frac{\epsilon}{2\Phi^2}\left[\frac{\stackrel{\ast}{\Phi} \stackrel{\ast}{g}_{\alpha \beta}}{\Phi} - \stackrel{\ast \ast}{g}_{\alpha \beta} + g^{\lambda\mu}\stackrel{\ast}{g}_{\alpha\lambda}\stackrel{\ast}{g}_{\beta\mu} - \frac{1}{2}g^{\mu\nu}\stackrel{\ast}{g}_{\mu\nu}\stackrel{\ast}{g}_{\alpha\beta} + \frac{1}{4}g_{\alpha\beta}\left(\stackrel{\ast}{g}^{\mu\nu}\stackrel{\ast}{g}_{\mu\nu} + (g^{\mu\nu}\stackrel{\ast}{g}_{\mu\nu})^2\right)\right],\nonumber \\
 &+& \frac{\Phi_{\alpha;\beta}}{\Phi}.
\end{eqnarray}
The energy-momentum tensor in $5D$ is usually taken as
\begin{equation}
\label{ETM in 5D}
{^{(5)}T}_{AB} = \Lambda_{(5)}\gamma_{AB},
\end{equation}
where $\Lambda_{(5)}$ is the cosmological constant in $5D$. Therefore,
the first term on the r.h.s. of (\ref{4D Einstein with T and K}) yields
\begin{equation}
T_{\mu\nu}^{(\Lambda)} = \frac{1}{2}k_{(5)}^2\Lambda_{(5)}g_{\mu\nu} \equiv \Lambda g_{\mu\nu},
\end{equation}
which defines an effective cosmological constant in $4D$ as $\Lambda = k_{(5)}^2\Lambda_{(5)}/2$.

As a consequence of the contracted Bianchi identities in $4D$, ${^{(4)}G}^{\mu}_{\nu; \mu} = 0$,  and (\ref{ETM in 5D}), the effective energy-momentum tensor satisfies the standard general relativity conservation equations, viz., 
\begin{equation}
D^{\mu}T_{\mu\nu}^{(eff)} = 0.
\end{equation}
If in the bulk there were scalar and/or other fields, then this would be no longer true, in general.

In addition, the  field equations in $5D$ reduce to
\begin{equation}
\label{field equations for Lambda neq zero}
R_{AB} = - \frac{2}{3}k_{(5)}^2 \Lambda_{(5)}\gamma_{AB}.
\end{equation}

\section{Spherically symmetric spacetime}
 
In this section we will discuss the extra symmetry of the field equations in $5D$,  in the case of spatial spherical symmetry. Thus, in what follows we will consider the metric
\begin{equation}
\label{the basic metric}
dS^2 = e^{\nu}dt^2 - e^{\lambda}dr^2 - R^2(d\theta^2 + \sin^2\theta d\phi^2) + \epsilon e^{\mu} dy^2,
\end{equation}
where the metric coefficients $\nu$, $\lambda$, $R$ and $\mu$ are allowed to be functions of the ``extra" coordinate $y$.

For the sake of clarity we provide here the non-zero components of the Ricci tensor. In the usual notation $\dot{f} \equiv \partial f/\partial t$, ${f'} \equiv \partial f/\partial r$ and $\stackrel{\ast}{f} \equiv \partial f/\partial y$, they are
\begin{eqnarray}
\label{R00}
R_{00} &=& \left( \frac{\dot{\nu}\dot{\lambda}}{4} + \frac{\dot{\nu}\dot{\mu}}{4} + \frac{\dot{\nu}\dot{R}}{R} - \frac{\ddot{\lambda}}{2} - \frac{\ddot{\mu}}{2} - 2\frac{\ddot{R}}{R}  - 
\frac{{\dot{\lambda}}^2}{4} - \frac{{\dot{\mu}}^2}{4}\right)\nonumber \\
 &+&  e^{\nu - \lambda}
\left(\frac{\nu''}{2} +  \frac{{\nu'}^2}{4} - \frac{\nu' \lambda'}{4} + \frac{\nu' \mu'}{4} + \frac{\nu' R'}{R}\right)
 + \epsilon e^{\nu - \mu}\left(\frac{\stackrel{\ast}{\nu} \stackrel{\ast}{\mu}}{4} - \frac{\stackrel{\ast \ast}{\nu}}{2} - \frac{{\stackrel{\ast}{\nu}}^2}{4} - \frac{\stackrel{\ast}{\nu} \stackrel{\ast}{\lambda}}{4}  - \frac{\stackrel{\ast}{\nu}\stackrel{\ast}{ R}}{R}\right),
\end{eqnarray}

\begin{eqnarray}
\label{R11}
R_{11} &=& \left(\frac{{\lambda}'{\mu}'}{4} + \frac{{\lambda}'{\nu}'}{4}
 + \frac{{\lambda}'{R}'}{R} - \frac{{\mu}''}{2} - \frac{{\nu}''}{2} - 2\frac{{R}''}{R}  - 
\frac{{\mu}'^2}{4} - \frac{{{\nu}'}^2}{4}\right)\nonumber \\
 &+& \epsilon e^{\lambda - \mu}\left(\frac{\stackrel{\ast\ast}{\lambda}}{2} +  \frac{\stackrel{\ast}{\lambda}^2}{4} - \frac{\stackrel{\ast}{\lambda} \stackrel{\ast}{\mu}}{4} + \frac{\stackrel{\ast}{\lambda} \stackrel{\ast}{\nu}}{4}  + \frac{\stackrel{\ast}{\lambda} \stackrel{\ast}{R}}{R}\right)
+  e^{\lambda - \nu}\left(\frac{\ddot{\lambda}}{2} + \frac{{\dot{\lambda}}^2}{4} - \frac{\dot{\lambda} \dot{\nu}}{4} +  \frac{\dot{\lambda} \dot{\mu}}{4} + \frac{\dot{\lambda}\dot{ R}}{R}\right), 
\end{eqnarray}
\begin{eqnarray}
\label{R22}
R_{22} &=& 1 + R^2 e^{- \nu}\left[\frac{\dot{R}^2}{R^2} + \frac{\ddot{R}}{R} - \frac{\dot{R}}{2R}(\dot{\nu} - \dot{\lambda} -  \dot{\mu})\right]\nonumber \\
&-& R^2 e^{- \lambda}\left[\frac{R'^2}{R^2} + \frac{R''}{R} + \frac{R'}{2R}(\nu' - \lambda'  + \mu')\right] 
+ \epsilon R^2 e^{- \mu}\left[\frac{\stackrel{\ast}{R}^2}{R^2} + \frac{\stackrel{\ast\ast}{R}}{R} + \frac{\stackrel{\ast}{R}}{2R}(\stackrel{\ast}{\nu} + \stackrel{\ast}{\lambda} - \stackrel{\ast}{\mu} )\right] 
\end{eqnarray}
\begin{equation}
R_{33} = \sin^2\theta R_{22},
\end{equation}

\begin{eqnarray}
\label{R44}
R_{44} &=&\left( \frac{\stackrel{\ast}{\mu}\stackrel{\ast}{\lambda}}{4} + \frac{\stackrel{\ast}{\mu}\stackrel{\ast}{\nu}}{4} + \frac{\stackrel{\ast}{\mu}\stackrel{\ast }{R}}{R} - \frac{\stackrel{\ast \ast}{\lambda}}{2} - \frac{\stackrel{\ast \ast}{\nu}}{2} - 2\frac{\stackrel{\ast \ast}{R}}{R} - 
\frac{\stackrel{\ast}{\lambda}^2}{4} - \frac{{\stackrel{\ast}{\nu}}^2}{4}\right)\nonumber \\
 &+& \epsilon e^{\mu - \lambda}\left(\frac{\mu''}{2} +  \frac{{\mu'}^2}{4} - \frac{\mu' \lambda'}{4} + \frac{\mu' \nu'}{4}  + \frac{\mu' R'}{R}\right)
+ \epsilon e^{\mu - \nu}\left(\frac{\dot{\mu} \dot{\nu}}{4} - \frac{\ddot{\mu}}{2} - \frac{{\dot{\mu}}^2}{4}  -  \frac{\dot{\mu} \dot{\lambda}}{4} - \frac{\dot{\mu}\dot{ R}}{R}\right), 
\end{eqnarray}
and 

\begin{eqnarray}
\label{R01, R41 and R04}
R_{01} &=&  \frac{\nu' \dot{\mu}}{4} + \frac{\dot{\lambda} \mu'}{4} 
+ \frac{\dot{\lambda} R'}{R} + \frac{\nu'\dot{R}}{R} - \frac{\dot{\mu'}}{2} - \frac{\dot{\mu} \mu'}{4}  - 2 \frac{\dot{R'}}{R},\nonumber \\
R_{41} &=&  \frac{\mu' \stackrel{\ast}{\nu}}{4} + \frac{\stackrel{\ast}{\lambda} \nu'}{4} 
+ \frac{\stackrel{\ast}{\lambda} R'}{R} + \frac{\mu'\stackrel{\ast}{R}}{R} - \frac{\stackrel{\ast}{\nu'}}{2} - \frac{\stackrel{\ast}{\nu} \nu'}{4} - 2 \frac{\stackrel{\ast}{R'}}{R},\nonumber \\
R_{04} &=&  \frac{\stackrel{\ast}{\nu} \dot{\lambda}}{4} + \frac{\dot{\mu} \stackrel{\ast}{\lambda}}{4} 
+ \frac{\dot{\mu} \stackrel{\ast}{R}}{R} + \frac{\stackrel{\ast}{\nu}\dot{R}}{R} - \frac{\stackrel{\ast}{\dot{\lambda}}}{2} - \frac{\dot{\lambda} \stackrel{\ast}{\lambda}}{4} - 2 \frac{\stackrel{\ast}{\dot{R}}}{R}.
\end{eqnarray}

Let us notice some properties of the above equations:

\begin{enumerate}
\item If we assume that the  metric coefficients are independent on $y$, then  the extra dimension can be either spacelike or timelike, without affecting the effective matter distribution in $4D$.

Indeed, a simple examination of (\ref{R00})-(\ref{R01, R41 and R04}) with $\stackrel{\ast}{\nu} = \stackrel{\ast}{\lambda} =  \stackrel{\ast}{\mu}  =  \stackrel{\ast}{R} = 0$, indicates that $\epsilon$, the signature of the extra dimension, enters nowhere, except in $R_{44}$, which becomes $R_{44} = \epsilon \times$(some function of $t$ and $r$). But $\gamma_{44} = \epsilon \Phi^2$. Then, from (\ref{EMT in STM}) and (\ref{field equations for Lambda neq zero}) it follows that both the field equations and the effective $4D$ matter are invariant with respect to the change of sign of $\epsilon$. 

\item If $y$ is spacelike $(\epsilon = -1)$, then we find that the field equations in $5D$ are invariant under the transformation
\begin{equation}
\label{If y is spacelike}
r \longleftrightarrow y,\;\;\;\lambda \longleftrightarrow \mu, \;\;\;\frac{\partial}{\partial r}\longleftrightarrow \frac{\partial}{\partial y}.
\end{equation}
Indeed, $R_{00}$, $R_{22}$ and $R_{14}$ remain invariant and $R_{01} \longleftrightarrow R_{04}$, $R_{11} \longleftrightarrow R_{44}$. Consequently, if  (\ref{the basic metric}) with $\epsilon = -1$ is a solution of the field equation, then 
\begin{equation}
\label{The metric, for epsilon = -1}
dS^2 = e^{\nu}dt^2 - e^{\mu}dr^2 - R^2(d\theta^2 + \sin^2\theta d\phi^2) - e^{\lambda} dy^2,
\end{equation}
is also a solution. We note from (\ref{EMT in STM}) that the effective matter is {\it not} invariant under such transformation, which means that (\ref{the basic metric}) and (\ref{The metric, for epsilon = -1}) lead to different scenarios in $4D$.

\item If $y$ is timelike $(\epsilon = + 1)$, we find that the field equations are invariant with respect to the transformation
\begin{equation}
\label{If y is timelike}
t \longleftrightarrow y,\;\;\;\nu \longleftrightarrow \mu, \;\;\;\frac{\partial}{\partial t}\longleftrightarrow \frac{\partial}{\partial y}.
\end{equation}
In this case $R_{11}$, $R_{22}$ and $R_{04}$ are invariant, while $R_{00} \longleftrightarrow R_{44}$ and $R_{01} \longleftrightarrow R_{41}$. Therefore, from (\ref{the basic metric}) with $\epsilon = 1$, it follows that 
\begin{equation}
\label{The metric, for epsilon = 1}
dS^2 = e^{\mu}dt^2 - e^{\lambda}dr^2 - R^2(d\theta^2 + \sin^2\theta d\phi^2) +  e^{\nu} dy^2,
\end{equation}
also satisfies the field equations. This metric and (\ref{the basic metric}) with time like extra dimension yield different scenarios in $4D$. 
\end{enumerate}

\section{$5D$ metrics with ``no" dependence on the extra dimension}

This is an important case because in $5D$ there are a number of solutions to the field equations obtained under the assumption that 
 $\stackrel{\ast}{\nu} = \stackrel{\ast}{\lambda} =  \stackrel{\ast}{\mu}  =  \stackrel{\ast}{R} = 0$. 

Usually, the extra coordinate $y$ is assumed to be spacelike, nevertheless  this is not a requirement of the field equations. Rather, we have seen that 
 \begin{equation}
\label{The metric no dependence of y}
dS_{(\pm)}^2 = e^{\nu(t, r)}dt^2 - e^{\lambda(t, r)}dr^2 - R^2(t, r)(d\theta^2 + \sin^2\theta d\phi^2)  + \epsilon e^{\mu(t, r)} dy^2,
\end{equation}
solves the field equations for both signatures $(\epsilon  = \pm 1)$, without affecting $4D$.
Therefore, there are two ``other" solutions associated with (\ref{The metric no dependence of y}),  depending on whether we choose $\epsilon = -1$ or $\epsilon = + 1 $. These are  

\begin{equation}
\label{The metric no dependence of y and epsilon = -1}
dS_{(-)}^2 = e^{\nu(t, y)}dt^2 - e^{\mu(t, y)}dr^2 - R^2(t, y)(d\theta^2 + \sin^2\theta d\phi^2)  -  e^{\lambda(t, y)} dy^2,
\end{equation}
 and  
\begin{equation}
\label{The metric no dependence of y and epsilon = 1}
dS_{(+)}^2 = e^{\mu(r, y)}dt^2 - e^{\lambda(r, y)}dr^2 - R^2(r, y)(d\theta^2 + \sin^2\theta d\phi^2)  +  e^{\nu(r, y)} dy^2,
\end{equation}
for $\epsilon = -1$ and $\epsilon = + 1$, respectively. 
Thus, we have started from a metric that does not depend on the extra dimension and finished with two metrics that do depend on it. Clearly, the property of being or not dependent of the extra dimension depends on how  we define it.

\paragraph{Interpretation in $4D$:}

For the four-dimensional interpretation of five-dimensional metrics we identify our 
spacetime with a  hypersurface orthogonal 
to the extra dimension, located at some value of $y$. 

Thus, from  (\ref{The metric no dependence of y}), (\ref{The metric no dependence of y and epsilon = -1}) and (\ref{The metric no dependence of y and epsilon = 1}) we find three different scenarios in $4D$, viz.,
\begin{equation}
\label{The 4D metric no dependence of y}
ds^2 = e^{\nu(t, r)}dt^2 - e^{\lambda(t, r)}dr^2 - R^2(t, r)(d\theta^2 + \sin^2\theta d\phi^2),
\end{equation}
\begin{equation}
\label{The 4D metric no dependence of y and epsilon = -1}
ds^2 = e^{\nu(t)}dt^2 - e^{\mu(t)}dr^2 - R^2(t)(d\theta^2 + \sin^2\theta d\phi^2),
\end{equation}
and
\begin{equation}
\label{The 4D metric no dependence of y and epsilon = 1}
ds^2 = e^{\mu(r)}dt^2 - e^{\lambda(r)}dr^2 - R^2(r)(d\theta^2 + \sin^2\theta d\phi^2).
\end{equation}
The first scenario (\ref{The 4D metric no dependence of y}) represents a non-static and spatially non-uniform spherical distribution of matter. The second one (\ref{The 4D metric no dependence of y and epsilon = -1}) is a cosmological metric of the Kantowski-Sachs type. The third scenario (\ref{The 4D metric no dependence of y and epsilon = 1}) corresponds to some static spherical distribution of matter. 

\subsection{Time-depending solutions}

In order to illustrate the above discussion let us consider the $5D$ metric

\begin{equation}
\label{the solution}
dS^2_{(\pm)} = B^2 \left[\frac{3r^2}{\alpha^2}dt^2 - t^2 dr^2 - \frac{t^2 
r^2}{(3 - \alpha^2)}(d\theta^2 + \sin^2 \theta d\phi^2)\right] \pm C^2 
r^{2(\alpha + 1)}t^{2(\alpha + 3)/\alpha}dy^2,
\end{equation}
which is a solution of the $5D$ field equations (\ref{field equations for Lambda neq zero}) with $\Lambda_{(5)} = 0$. Here $\alpha$ is a dimensionless 
parameter in the range $0 < \alpha^2 < 3$, whereas $B$ and $C$ are arbitrary 
constants with the dimensions $L^{- 1}$ and $L^{- (\alpha^2 + 2\alpha + 
3)/\alpha}$, respectively.

\medskip

The metric (\ref{the solution})  works for both signatures. Let us discuss the different possible scenarios.

\paragraph{$\epsilon = \pm 1$:}The induced metric in $4D$ 
can be interpreted as
a spherically symmetric dissipative distribution of matter, with heat flux, whose effective density and 
pressure are nonstatic, nonuniform, and satisfy the equation of state
of radiation. This interpretation is  not affected whatsoever by the signature of the extra 
dimension \cite{JPdeLgr-qc/0402046}.

\medskip

\paragraph{$\epsilon = - 1$:} If we take $y$ as a spacelike coordinate, and choose $3B^2 = \alpha^2$, then  (\ref{the solution}) yields\footnote{We note that, for 
this choice of the constants, the Kantowski-Sachs line element (\ref{the solution with epsilon = -1}) looks very similar to the standard FRW-flat cosmological model embedded in $5D$. Namely, $d{\cal S}^2 = y^2 dt^2 - A^2 t^{2/\alpha}y^{2/(1 - \alpha)}[dr^2 + r^2(d\theta^2 + \sin^2\theta d\phi^2)] - \alpha^2(1- \alpha)^{-2} t^2 dy^2$ \cite{JPdeL 1}}

\begin{equation}
\label{the solution with epsilon = -1}
dS^2_{(-)} = y^2dt^2 - \bar{C}^2y^{2(\alpha + 1)}t^{2(\alpha + 3)/\alpha}dr^2
- \frac{\alpha^2t^2y^2}{3(3 - \alpha^2)}(d\theta^2 + \sin^2\theta d\phi^2) - \frac{\alpha^2 t^2}{3}dy^2.
\end{equation}
In $4D$ this solution corresponds to  a cosmological model of Kantowski-Sachs type. These cosmologies have extensively been studied by the present author; they can be relevant to describe ``bubbles" of new phases, in phase transitions \cite{JPdeL1}-\cite{JPdeL2}. In particular, the $4D$ part of  
(\ref{the solution with epsilon = -1}) is the {\it only} Kantowski-Sachs universe that has the property of self-similarity. It represents a homogeneous  universe which is expanding with shear (rather than shear-free as in FRW cosmologies). The expansion $\Theta$ and shear $\sigma$ are given by \cite{JPdeL2}
\begin{equation}
\Theta = \frac{3(\alpha + 1)}{\alpha t}, \;\;\;\sigma = \frac{\sqrt{3}}{\alpha t}
\end{equation}  
We note that in (\ref{the solution with epsilon = -1}) the coordinate $y$ is dimensionless and 
$\bar{C} \sim L^{- (\alpha + 3)/\alpha}$.

 \paragraph{$\epsilon = + 1$:} If we take $y$ as a timelike coordinate, choose $B^2 = (3 - \alpha^3)$ and denote $C = r_{0}^{- (\alpha + 1)}$, then $y$ becomes dimensionless and (\ref{the solution}) yields 

\begin{equation}
\label{the solution with epsilon = +1}
dS^2_{(+)} = \left(\frac{r}{r_{0}}\right)^{2(\alpha + 1)}y^{2(\alpha + 3)/\alpha}dt^2 - (3 - \alpha^2)y^2dr^2 - y^2r^2(d\theta^2 + \sin^2\theta d\phi^2) + 3(3 \alpha ^{- 2} - 1) r^2 dy^2. 
\end{equation}
This solution was extensively studied by Billyard and Wesson \cite{Billyard and Wesson}. In $4D$, it represents  a spherical static cloud of matter with  density profiles similar to those of cluster of galaxies. 

There are many other spherical solutions of the five-dimensional field equations that 
depend on $t$ and $r$, but not on $y$ \cite{WLPdeL}-\cite{Kokarev}. All of them were obtained under the assumption of a spacelike extra dimension. Therefore, using the symmetries (\ref{If y is spacelike}) and (\ref{If y is timelike}), we can use them to generate new solutions of the type discussed above.

\subsection{Static solutions}

In the case of spherical symmetry in ordinary $3D$ space, there is only {\it one} family of  exact solutions of (\ref{field equations for Lambda neq zero}), with $\Lambda_{(5)} = 0$, which are (i) static and (ii) independent of the fifth coordinate; if one relaxes any of these conditions, then there are many solutions in $5D$ vacuum, with spherical $3$-space. 

This solution has been rediscovered many times and is known in the literature under 
different names, viz., as  Kramer's solution \cite{Kramer}
\medskip

\begin{equation}
\label{Kramer sol}
dS^2 = \left(1 - \frac{r_{g}}{r}\right)^{(A - B)}dt^2 - \left(1 - \frac{r_{g}}{r}\right)^{- (A + B)}dr^2 - r^2\left(1 - \frac{r_{g}}{r}\right)^{(1 - A - B)}d\Omega^2 - \left(1 - \frac{r_{g}}{r}\right)^{2B}dy^2,
\end{equation}
with $A^2 + 3B^2 = 1$ and $r_{g} = const$, as the  Davidson-Owen class of solutions \cite{Davidson Owen}
\medskip
\begin{equation}
\label{Davidson and Owen solution}
dS^2 = \left(\frac{ar - 1}{ar + 1}\right)^{2\sigma k}dt^2 - \frac{1}{a^4r^4}\frac{(ar + 1)^{2[\sigma(k - 1) + 1]}}{(ar - 1)^{2[\sigma(k - 1) - 1]}}[dr^2 + r^2d\Omega^2]
- \left(\frac{ar + 1}{ar -1}\right)^{2\sigma}dy^2,
\end{equation}
where $\sigma^2(k^2 - k + 1) = 1$ and $a = const.$, and, although in another context, as Gross and Perry solutions \cite{Gross Perry}. These solutions were discovered under the assumption of a spacelike extra coordinate, but they also work for a timelike extra dimension.

Distinct  forms of the solution are useful in different contexts, which include the generation of  new solutions, the description of extended spherical objects in $4D$ called solitons, and the study of geodesics in $5D$.

\subsubsection{New solutions}
In the above solutions the extra coordinate is assumed to be spacelike. However, according to our discussion, a ``new" solution arises from the symmetries of the field equations. Using (\ref{If y is timelike}) in (\ref{Kramer sol}) we obtain  
 \begin{equation}
\label{Kramer sol with two times}
dS^2 = \left(1 - \frac{r_{g}}{r}\right)^{2B}dt^2   - \left(1 - \frac{r_{g}}{r}\right)^{- (A + B)}dr^2 - r^2\left(1 - \frac{r_{g}}{r}\right)^{(1 - A - B)}d\Omega^2 +  \left(1 - \frac{r_{g}}{r}\right)^{(A - B)}dy^2,
\end{equation}
and in the and Davidson-Owen form 
\begin{equation}
\label{Davidson and Owen solution with two times}
dS^2 = \left(\frac{ar + 1}{ar -1}\right)^{2\sigma}dt^2  - \frac{1}{a^4r^4}\frac{(ar + 1)^{2[\sigma(k - 1) + 1]}}{(ar - 1)^{2[\sigma(k - 1) - 1]}}[dr^2 + r^2d\Omega^2]
+  \left(\frac{ar - 1}{ar + 1}\right)^{2\sigma k}dy^2  ,
\end{equation}

\medskip

The four-dimensional interpretation  of these metrics (at some $y = $const.) differs from the one for (\ref{Kramer sol}) and (\ref{Davidson and Owen solution}). 

\medskip

First, in (\ref{Kramer sol}) and (\ref{Kramer sol with two times}) the Schwarzschild solution is recovered in curvature coordinates for distinct values of $A$ and $B$. Namely, $A = 1$,   $B = 0$ for (\ref{Kramer sol}) and $A = B = 1/2$ for (\ref{Kramer sol with two times}). 
The original Davidson-Owen solution yields the Schwarzschild solution in isotropic coordinates in the {\it limit} $\sigma \rightarrow 0$, $k \rightarrow \infty$ and $\sigma k \rightarrow  1$, while in the new version 
(\ref{Davidson and Owen solution with two times}) this occurs for $\sigma = - 1$ and $k = 0$. 

Second, the effective energy-momentum tensor in $4D$ calculated from (\ref{Kramer sol with two times}) is distinct from the one calculated from (\ref{Kramer sol}), similarly for (\ref{Davidson and Owen solution with two times}) and (\ref{Davidson and Owen solution}), respectively. Indeed, from (\ref{EMT in STM}) it follows that
\begin {equation}
\label{EMT for solutions with two times }
8 \pi GT_{\mu\nu}^{(eff)} =   \frac{\Phi_{\alpha;\beta}}{\Phi},
\end{equation}
where $\epsilon \Phi^2$, is the metric coefficient in front of $dy^2$, which is different in  (\ref{Kramer sol}), (\ref{Davidson and Owen solution}) and (\ref{Kramer sol with two times}),(\ref{Davidson and Owen solution with two times}), respectively.

The soliton solutions (\ref{Kramer sol})-(\ref{Davidson and Owen solution}) play a central role in the discussion of  
many important observational problems, which include the classical tests of relativity, as well as the geodesic precession of a gyroscope and possible departures from the equivalence principle \cite{Wesson book}. Therefore, a detailed study of metrics (\ref{Kramer sol with two times}) and (\ref{Davidson and Owen solution with two times}) is important. It should allow us to understand  the influence of an extra timelike coordinate on ordinary four-dimensional physics. Such investigation is out of the scope of the present paper.

\section{$5D$ metrics with explicit dependence on the extra dimension and $\epsilon = \pm 1$}

The five-dimensional  metrics  (\ref{the solution with epsilon = -1}) and (\ref{the solution with epsilon = +1}) show  explicit dependence on the extra dimension, but the signature is restricted to be either spacelike or timelike, respectively. On the other hand, there are a number of solutions of the $5D$ field equations that depend on the extra dimension and  allow both signatures.

For solutions of this kind the extra symmetry produces a large ``family" of different scenarios in $4D$. As an illustration let us consider the metric
\begin{equation}
\label{static sol for epsilon pm 1}
dS_{(\pm)}^2 = \frac{A^2\alpha^2 y^2}{(\alpha y^2 + \beta)}dt^2 - \frac{(\alpha y^2 + \beta)}{(ar^2 + b)^2}[dr^2 + r^2d\Omega^2] + \epsilon  dy^2,
\end{equation}
which is an exact solution of the field equations in $5D$, provided\footnote{Under the transformation $\bar{r} = Br$, $\bar{y} = C y$; 
 $\bar{\alpha} = \alpha C^2$ and $\bar{a} = aB^2$, therefore $\epsilon \bar{\alpha} (B/C)^2 = - 4\bar{a}b.$}  $\alpha \epsilon = - 4 a b$. 
Here $A$ is an arbitrary constant with dimensions of $L$, $\alpha$ and $a$ are constants with dimensions $L^{- 2}$, while $\beta$ and $b$ are dimensionless constants. The properties of this solution were discussed in Ref. \cite{JPdeL Wesson}. In $4D$ it represents a  matter distribution which satisfies the equation of state\footnote{It is interesting to note that this equation of state appears in very different contexts; in an alternative derivation of properties of matter from $5D$ geometry by Davidson and Owen\cite{Davidson Owen}; in discussions of cosmic strings by Gott and Rees \cite{Goot and Rees} and Kolb \cite{Kolb}; in certain sources (called ``limiting configurations") for the Reissner-Norsdstr\"{o}m field by the present author \cite{Present author} and as the only equation of state consistent with the existence of quantum zero-point fields by 
Wesson \cite{W1}-\cite{W2}} $\rho = - 3p$, where $8\pi G \rho = 12 ab(\alpha y^2 + \beta)^{- 1}$.   

We have several choices for the extra dimension here. If $\epsilon = - 1$, from (\ref{If y is spacelike}) we get 
\begin{equation}
\label{static sol for epsilon - 1}
dS_{(-)}^2 = \frac{A^2 \alpha^2 r^2}{(\alpha r^2 + \beta)}dt^2 - dr^2 - 
\frac{y^2(\alpha r^2 + \beta)}{(ay^2 + b)^2}d\Omega^2  - \frac{(\alpha r^2 + \beta)}{(ay^2 + b)^2}dy^2.
\end{equation}
If $\epsilon = 1$, from (\ref{If y is timelike}) we get 
\begin{equation}
\label{static sol for epsilon = 1}
dS_{(+)}^2 = dt^2    - \frac{(\alpha t^2 + \beta)}{(ar^2 + b)^2}[dr^2 + r^2d\Omega^2] 
+ \frac{A^2 \alpha^2 t^2}{(\alpha t^2 + \beta)}dy^2.
\end{equation}
This metric depends on $t$ and $r$ only. Therefore, it should solve the equations for both 
signatures, i.e.,
\begin{equation}
\label{r-t sol for both signatures}
dS_{(-)}^2 = dt^2    - \frac{(\alpha t^2 + \beta)}{(ar^2 + b)^2}[dr^2 + r^2d\Omega^2] 
- \frac{A^2 \alpha^2 t^2}{(\alpha t^2 + \beta)}dy^2.
\end{equation}
is also a solution. From here and (\ref{If y is spacelike}) we get  the Kantowski-Sachs metric
\begin{equation}
\label{r-t sol transforms into KS metric}
dS_{(-)}^2 = dt^2 - \frac{A^2 \alpha^2 t^2}{(\alpha t^2 + \beta)}dr^2 - \frac{y^2(\alpha t^2 + \beta)}{(ay^2 + b)}d\Omega^2 -\frac{(\alpha t^2 + \beta)}{(\alpha y^2 + b)^2}dy^2.
\end{equation}
Despite the similarity between the above line elements, each represents a different scenario in $4D$, which is obtained at some $y = const.$. But all of them represent the same five-dimensional solution (\ref{static sol for epsilon pm 1}).

\section{Summary and concluding remarks}

The field equations in $5D$, with spatial spherical symmetry, have an additional symmetry that leave the equations invariant. This symmetry reflects the fact that there are many ways of  producing, or embedding,  a $4D$ spacetime in a given five-dimensional manifold, while satisfying the field equations.

From a practical viewpoint this symmetry allows us to generate diverse physical scenarios in $4D$ from a given solution in $5D$. From a theoretical point of view it unifies in five-dimensions a number of, otherwise detached, dynamical models in $4D$.

Thus, our work provides a new perspective and new challenges. The new perspective is that different astrophysical and cosmological scenarios in $4D$ might correspond to  the  same higher 
dimensional system, with different choices of signature and extra coordinate.

The new challenge is to understand in more detail the relationship between apparently different $4D$ scenarios, belonging to the same five-dimensional family, as well as the nature and effects of a timelike extra dimension.    

We would like to finish this paper with the following comments. 
\begin{enumerate}
\item We have never seen in the literature the metric (\ref{Kramer sol with two times}) nor (\ref{Davidson and Owen solution with two times}). To our best knowledge they have never been used in the discussion of the classical and other tests in $5D$. Certainly, the study of these metrics in this context would give a better understanding of the subject.

\item It can be shown that the solution (\ref{static sol for epsilon pm 1}) also satisfies the $5D$ field equations (with $\Lambda_{(5)} = 0$) with the constants $a$, $b$ replaced by functions of $y$. Therefore, applying the transformations  (\ref{If y is spacelike}) and (\ref{If y is timelike})  to it,   a large family of other $5D$ solutions will emerge   that generate a variety of physical models in $4D$.   We hope to investigate these models in future work.

\item In both, compactified and non-compactified theories, the extra 
dimensions  are  usually assumed  to be spacelike. However, there 
is no {\em a priori} reason why  extra dimensions cannot be timelike. 
As a matter of fact, the consideration of extra timelike  dimensions 
in physics has a long and distinguished history \cite{Dirac}, \cite{Kastrup}, 
\cite{Salam}, \cite{Sakharov} and currently it is a subject of considerable 
interest. For a more detailed discussion and references see for example \cite{JPdeLgr-qc/0212058}.

\end{enumerate}

\end{document}